\documentclass[aps,prc,preprint,tightenlines,groupedaddress,nofootinbib,
showpacs,preprintnumbers,amsmath,amssymb,superscriptaddress]{revtex4}
\usepackage{graphicx}
\usepackage{dcolumn}
\usepackage{mathrsfs}
\usepackage{bm}

\begin{document}
\title{Impact of the symmetry energy on the outer crust of 
       non-accreting neutron stars}  
\author{X. Roca-Maza}
\affiliation{Departament d'Estructura i Constituents 
             de la Mat\`eria, Facultat de F\'{\i}sica,\\    
             Universitat de Barcelona, Diagonal {\sl 647}, 
             {\sl 08028} Barcelona, Spain}
\author{J. Piekarewicz}
\affiliation{Department of Physics, Florida State 
             University, Tallahassee, FL 32306}
\date{\today} 

\begin{abstract}
 The composition and equation of state of the outer crust 
 of non-accreting neutron stars is computed using accurate 
 nuclear mass tables. The main goal of the present study is 
 to understand the impact of the symmetry energy on the 
 structure of the outer crust. First, a simple {\sl ``toy 
 model''} is developed to illustrate the competition between 
 the electronic density and the symmetry energy. Then, 
 realistic mass tables are used to show that models with a 
 stiff symmetry energy --- those that generate large neutron 
 skins for heavy nuclei --- predict a sequence of nuclei that 
 are more neutron-rich than their softer counterparts. This 
 result may be phrased in the form of a correlation: 
 {\emph{the larger the neutron skin of ${}^{208}${\rm Pb}, 
 the more exotic the composition of the outer crust.}}  
\end{abstract}
\maketitle 

\section{Introduction}
\label{Introduction}

\begin{figure}[h]
 \includegraphics[width=3in]{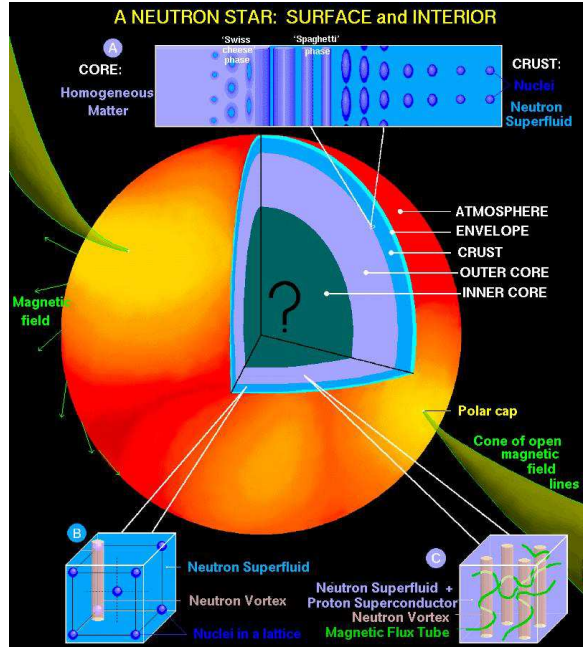}
 \vspace{-0.2cm}
 \caption{(color online)
           Rendition of the assumed structure and phases of
           a neutron star (courtesy of Dany Page).}
 \label{Fig0}
\end{figure}

Neutron stars are gold mines for the study of nuclear systems under
extreme conditions of density and isospin
asymmetry~\cite{Lattimer:2000nx, Lattimer:2004pg}. Spanning many
orders of magnitude in density, neutron stars display exotic phases
that cannot be realized under normal laboratory conditions. While the
most common perception of a neutron star is that of a uniform mantle
of neutrons packed to densities that may exceed that of normal nuclei
by up to an order of magnitude, the reality is far different and much
more interesting.  First, although the uniform liquid mantle (also
known as the outer core) is indeed composed mostly of neutrons, a
small fraction of protons and an equal number of charged leptons ({\it
i.e.,} electrons and perhaps even muons) must be present to maintain
beta equilibrium. The precise proton fraction in the neutron star is
controlled by the {\emph{symmetry energy}}, a quantity that imposes a
penalty on the system as it departs from the isospin symmetric limit
of equal number of neutrons and protons. Second, at densities that are
below nuclear matter saturation density the uniform phase becomes
unstable against density fluctuations. This non-uniform region of the
neutron star constitutes the {\emph{crust}}, which itself is divided
into an inner and an outer region (see Fig.~\ref{Fig0}).  In the outer
crust --- the main focus of the present study --- the system is
organized into a Coulomb lattice of neutron-rich nuclei embedded in a
uniform electron gas~\cite{Baym:1971pw}.  As the density increases,
nuclei become progressively more neutron rich until the neutron drip
region is reached; this region defines the boundary between the outer
and the inner crust. As in the case of the outer crust, the inner
crust also consists of a Coulomb lattice of neutron-rich nuclei
embedded in a uniform electron gas. Now, however, a uniform neutron
vapor permeates the system. As the density continues to increase
in the inner crust, the system is speculated to morph into a variety
of complex and exotic structures, such as spheres, cylinders, rods,
plates, {\it etc.} --- collectively known as {\emph{nuclear
pasta}}~\cite{Ravenhall:1983uh,Hashimoto:1984}. As the density
increases even further, uniformity is eventually restored at about one
third of normal nuclear matter saturation density.  Finally, at ultra
high densities it has been established that the ground state of
hadronic matter becomes a color superconductor in a
color-flavor-locked (CFL)
phase~\cite{Alford:1998mk,Rajagopal:2000wf}. It is unknown, however,
if the density at the core of a neutron star may reach the extreme
values required for the CFL phase to develop. Thus, other exotic
phases --- such as meson condensates, hyperonic matter, and/or quark
matter --- may be more likely to harbor the core of neutron
stars. Figure~\ref{Fig0} is believed to represent a plausible 
rendition of the structure of a neutron star.

As stated earlier, the main focus of our present study is the outer
crust of the neutron star. In particular, we are interested in
studying the sensitivity of the composition of the outer crust to the
model dependence of the symmetry energy. The outer crust comprises a
region spanning about seven orders of magnitude in density; from about
$10^{4}{\rm g/cm^{3}}$ up to a neutron-drip density of about
$4\times10^{11}{\rm g/cm^{3}}$~\cite{Baym:1971pw}. Although small
relative to nuclear matter saturation density ($2.5\times10^{14}{\rm
g/cm^{3}}$), at these densities the electrons (present to maintain
charge neutrality) are no longer bound to nuclei and move freely
throughout the crust. Moreover, at these low nuclear densities it is
energetically favorable for the nuclei to arrange themselves in a
crystalline lattice. At the lowest densities, the electronic
contribution is negligible so the Coulomb lattice is populated by
${}^{56}$Fe nuclei. However, as the density increases and the
electronic contribution becomes important, ${}^{56}$Fe ceases to be
the most energetically favorable nucleus. Instead, it becomes
energetically advantageous for the system to lower its electron
fraction by having the energetic electrons capture onto protons, with
the excess energy carried away by neutrinos. The resulting nuclear
lattice is now formed by nuclei having a slightly lower proton
fraction than ${}^{56}$Fe ({\it e.g.,} ${}^{62}$Ni). As the density
continues to increase, the nuclear system evolves into a Coulomb
lattice of progressively more neutron-rich nuclei until the critical
neutron-drip density is reached. The essential physics of the outer
crust is then nicely captured by a {\sl ``tug-of-war''} between an
electronic contribution and the nuclear symmetry energy, with the
former favoring neutron-rich nuclei while the latter favoring fairly
symmetric ones. The neutron-rich nuclei that populate the Coulomb
lattice in the outer crust are on average more dilute than their more
symmetric counterparts because of the development of a neutron
skin. As a result, these nuclei may become sensitive to the symmetry
energy below nuclear matter saturation density. However, whereas the
symmetry energy is relatively well known around saturation density,
its density dependence ({\it e.g.,} its slope) is poorly
constrained. This may affect the composition of the crust.

While some theoretical constraints are starting to
emerge~\cite{Schwenk:2005ka,Piekarewicz:2007dx}, the density
dependence of the symmetry energy remains largely unknown. Indicative
of this fact is that accurately-calibrated models of nuclear structure
(both relativistic and non-relativistic) that reproduce a variety of
ground-state properties across the periodic table differ significantly
in their predictions for the density dependence of the symmetry
energy. Yet these same models have been used to uncover a strong
correlation between the pressure of pure neutron matter at saturation
density and the neutron skin of heavy nuclei: the larger the pressure
the larger the neutron skin~\cite{Brown:2000, Furnstahl:2001un}. (Note
that the pressure of pure neutron matter equals that of the symmetry
energy at saturation density.) This fact may be illustrated using the
two accurately-calibrated models that will be employed throughout this
manuscript, namely, NL3~\cite{Lalazissis:1996rd, Lalazissis:1999} and
FSUGold~\cite{Todd-Rutel:2005fa}. Whereas NL3 predicts a pressure at
saturation density of $P_{0}\!\approx\!6~{\rm MeV/fm}^{3}$ and a
corresponding neutron skin in ${}^{208}$Pb of $R_{n}\!-\!R_{p}\!
\approx\!0.28$~fm, FSUGold predicts the significantly lower values of
$P_{0}\!\approx\!3~{\rm MeV/fm}^{3}$ and $R_{n}\!-\!R_{p}\!\approx\!
0.21$~fm, respectively. The upcoming {\emph{Parity Radius Experiment}}
(PREx) at the Jefferson Laboratory will provide a unique experimental
constraint on the density dependence of the symmetry by measuring the
skin thickness of $^{208}$Pb accurately and model independently via
parity-violating electron scattering~\cite{Horowitz:1999fk,
Michaels:2005}.  The correlation between the density dependence of the
symmetry energy and the neutron skin of heavy nuclei opened new
horizons in nuclear astrophysics. Novel correlations between the
neutron skin of ${}^{208}$Pb and a myriad of neutron-star observables
were developed as a result of the similar composition of the neutron
skin of a heavy nucleus and the inner-crust/outer-core of a neutron
star~\cite{Horowitz:2000xj,Horowitz:2001ya,Horowitz:2002mb,
Carriere:2002bx, Steiner:2004fi}. One particularly interesting
correlation of direct relevance to the crustal region is a {\sl
``data-to-data''} relation between the neutron skin of ${}^{208}$Pb 
and the crust-to-core transition density~\cite{Horowitz:2000xj}.

The recent observation of intense pulses of energetic gamma rays
followed by fainter periodic signals emitted from highly magnetized
neutron stars (or {\sl {``magnetars''}}) are sure to provide an
additional new tool in the study of neutron-star
structure~\cite{Piro:2005jf,Strohmayer:2006py,Watts:2006ew,Watts:2006mr}.
The discovery of high-frequency oscillations in the tails of giant
flares from Soft Gamma Repeaters, {\it i.e.,} magnetars with magnetic
fields in excess of $10^{14}$
gauss~\cite{Thompson:1995gw,Kouveliotou:1998ze, Kouveliotou:2003tb},
are believed to be associated with seismic vibrations of the neutron
star crust. Early theoretical models that assume a
liquid-core/solid-crust interface suggest torsional shear oscillations
of the crust as the most likely modes of excitation in a magnetar. The
shear modulus of the crust acts as a restoring force for these modes
and such a structural property is highly sensitive to the composition
of the crust and, thereby, to the nuclear matter equation of
state. Indeed, the shear-mode oscillations depend strongly on the
neutron star mass, radius, and {\emph{crustal composition}} --- all
properties sensitive to equation of state~\cite{Piro:2005jf}.
Moreover, {\sl ratios} of frequencies with different nodal structures
may be used to determine the thickness of the crust, an observable
highly sensitive to the equation of state and particularly to the
density dependence of the symmetry energy~\cite{Strohmayer:2006py,
Watts:2006ew}. Hence, as techniques continue to improve, we expect
that neutron-star seismology will provide stringent limits on the 
equation of state of neutron-rich matter.

The manuscript has been organized as follows. The formalism required
to compute the composition and equation of state of the outer crust is
developed in Sec.~\ref{Formalism}. In Sec.~\ref{Results} we employ
several realistic nuclear-mass models to compute the structure of the
outer crust. While not nearly as comprehensive as the recent study
performed by Ruester, Hempel, and
Schaffner-Bielich~\cite{Ruester:2005fm}, ours include a simple {\sl
``toy model''} that provides critical insights into the role played by
the symmetry energy. Moreover, in the same section we illustrate the
impact of the density dependence of the symmetry energy on the
sequence of neutron-rich nuclei present in the outer crust. Our
results and conclusions are summarized in Sec.~\ref{Conclusions}.

\section{Formalism}
\label{Formalism}

In this section we develop the formalism necessary to understand the
composition and equation of state of the outer crust of a neutron
star. The formalism follows closely the seminal ideas introduced by
Baym, Pethick, and Sutherland back in 1971~\cite{Baym:1971pw}. For
more recent references that employ modern nuclear mass tables
see Refs.~\cite{Haensel:1989,Haensel:1994,Ruester:2005fm}. The central
question that one aims to answer is the following: what is the ground
state of cold, fully-catalyzed matter for densities between complete
ionization ($\rho\!\approx\!10^{4} {\rm g/cm^{3}}$) and {\sl ``neutron 
drip''} ($\rho\!\approx\!10^{11}{\rm g/cm^{3}}$)?  Since at
these densities uniform matter is unstable against cluster formation,
a Coulomb lattice of nuclei embedded in a uniform free Fermi gas of
electrons is formed. Thus, the composition of the outer crust is
determined by that nucleus (with neutron number $N$, proton number
$Z$, and baryon number $A\!=\!N\!+\!Z$) that minimizes --- for each
density --- the total energy per nucleon of the system. In the outer
crust ({\it i.e.,} before neutron drip) the {\sl energy per nucleon}
consists of three different contributions: nuclear, electronic, and
lattice. That is,
\begin{equation}
 \varepsilon(A,Z;n)=\varepsilon_{n}+\varepsilon_{e}+
                    \varepsilon_{\ell}\;,
 \label{EoverA}
\end{equation}
where the baryon density is denoted by $n\!\equiv\!A_{\rm total}/V$. The 
nuclear contribution to the total energy per nucleon is simple and 
independent of the density. It is given by
\begin{equation}
 \varepsilon_{n}(N,Z)\equiv\frac{M(N,Z)}{A}\;,  \;\;{\rm with}\;\;
  M(N,Z)=Nm_{n}+Zm_{p}-B(N,Z) \;.
 \label{EoverANucl}
\end{equation}
Here $M(N,Z)$ is the nuclear mass, $B(N,Z)$ is the corresponding 
binding energy, and $m_{n}$ and $m_{p}$ are neutron and proton 
masses, respectively.

The electronic contribution --- at least at the densities of 
interest ($\rho\!\gtrsim\!10^{4} {\rm g/cm^{3}}$) --- is modeled
as a degenerate free Fermi gas~\cite{Baym:1971pw}. That is,
\begin{equation}
 \varepsilon_{e}(A,Z;n)=\frac{\mathscr{E}_{e}}{n}
  =\frac{1}{n\pi^{2}}\int_{0}^{p_{{\rm F}e}}
   p^{2}\sqrt{p^{2}+m_{e}^{2}}\,dp \;,
 \label{EoverAElec}
\end{equation}
where $\mathscr{E}_{e}$, $m_{e}$, and $p_{{\rm F}e}$ are the electronic 
energy density, mass, and Fermi momentum, respectively. Note that the 
electronic Fermi momentum and baryon density are related as follows:
\begin{equation}
 p_{{\rm F}e} = \left(3\pi^2 n_{e}\right)^{1/3}
             = \left(3\pi^2 y n\right)^{1/3} 
               \equiv y^{1/3}p_{\rm F}\;,
 \label{pFermie}
\end{equation}
where the electron fraction $y\!\equiv\!Z/A$ has been defined. 
Moreover, for future convenience the following definition of 
the overall Fermi momentum has been introduced:
\begin{equation}
 p_{{\rm F}} = \left(3\pi^2 n\right)^{1/3} \;.
 \label{pFermi}
\end{equation}
As the integral in Eq.~(\ref{EoverAElec}) may be evaluated
analytically, the electronic contribution may be computed in
closed form. That is,
\begin{equation}
 \varepsilon_{e}(A,Z;n)=\frac{m_{e}^{4}}{8\pi^{2}n}
 \left[x_{\rm F}y_{\rm F}\Big(x_{\rm F}^{2}+y_{\rm F}^{2}\Big)
      -\ln(x_{\rm F}+y_{\rm F})\right]\;,
 \label{EoverAElec2}
\end{equation}
where dimensionless Fermi momentum and energy have been defined
as follows:
\begin{equation}
  x_{\rm F}\!\equiv\!\frac{p_{{\rm F}e}}{m_{e}} 
        \;\;{\rm and}\;\;
  y_{\rm F}\!\equiv\!\frac{\epsilon_{{\rm F}e}}{m_{e}}=
          \sqrt{1+x_{\rm F}^{2}}\;.
 \label{peFermi}
\end{equation}

We now discuss the last term in Eq.~(\ref{EoverA}).  Whereas the
Coulomb repulsion within the individual nuclei has been properly
included in Eq.~(\ref{EoverANucl}), the Coulomb repulsion among nuclei
as well as their interactions with the uniform electron background has
not. At the densities/temperatures of relevance to the outer crust,
namely, large enough for full ionization but small enough for the
Coulomb repulsion among nuclei to dominate over their kinetic energy
--- Wigner has shown (in the context of the electron gas) that the
system will crystallize into a body-centered-cubic
lattice~\cite{Wigner:1934, Wigner:1938,Fetter:1971}. The last term in
Eq.~(\ref{EoverA}) represents the lattice contribution to the energy
per particle. The calculation of the potential energy of the Coulomb
lattice is complicated. It consists of divergent contributions that
must be canceled as required by the overall charge neutrality of the
system. Fortunately, accurate numerical calculations for the electron
gas have been available for a long time~\cite{Coldwell:1960,Sholl:1967} 
and these results can be readily generalized to the present 
case~\cite{Baym:1971pw}. Indeed, the lattice energy per nucleus may be 
written as follows:
\begin{equation}
  \frac{E_{\ell}}{N_{c}} = -(1.81962)\frac{(Ze)^{2}}{a}
                       =  -(1.79186)\frac{(Ze)^{2}}{2r_{0}}\;.
 \label{EoverNLatt}
\end{equation}
where $N_{c}$ is the number of nuclei ({\it i.e.,} A-body clusters),
$a$ is the lattice constant, and $r_{0}$ is a length scale related
to the volume per nuclei. For the particular case of a 
body-centered-cubic lattice, these quantities are related in the 
following way:
\begin{equation}
  n_{c}a^{3}=\frac{N_{c}}{V}a^{3}=2 \quad{\rm or}\quad
  \left(\frac{a}{2r_{0}}\right)=\left(\frac{\pi}{3}\right)^{1/3} \;.
 \label{r0vsa}
\end{equation}
Using the fact that the number of nuclei is related to the total
baryon number of the system as 
\begin{equation}
  N_{c} = \frac{A_{\rm total}}{A} = \frac{N_{\rm total}}{N}
       = \frac{Z_{\rm total}}{Z}\;,
  \label{NClusters}
\end{equation}
the lattice contribution to the energy per baryon may be written
in closed form as follows:
\begin{equation}
 \varepsilon_{\ell}(A,Z;n)=
   -\frac{(1.79186)}{A^{4/3}}\frac{(Ze)^{2}}{2R_{0}}\;.
 \label{EoverALatt}
\end{equation}
Note that here $R_{0}$ refers to the length scale associated 
to the volume per {\sl baryon} ($r_{0}=A^{1/3}R_{0}$). That is,
\begin{equation}
  R_{0} = \left(\frac{3}{4\pi n}\right)^{1/3} 
       = \left(\frac{9\pi}{4}\right)^{1/3}p_{\rm F}^{-1} \;.
 \label{R0vsr0}
\end{equation}
Using these definitions, the lattice contribution becomes equal to
\begin{equation}
 \varepsilon_{\ell}(A,Z;n)=-
 C_{\ell} \frac{Z^{2}}{A^{4/3}}p^{}_{\rm F}
 \quad ({\rm with}\; C_{\ell}\!=\!3.40665\!\times\!10^{-3}) \;.
 \label{EoverALatt2}
\end{equation}
For completeness, the full expression for the energy per baryon is now 
displayed in terms of the individual {\sl nuclear, electronic,} plus 
{\sl lattice} contributions:
\begin{equation}
 \varepsilon(A,Z;n)=\frac{M(N,Z)}{A}+
  \frac{m_{e}^{4}}{8\pi^{2}n}
  \left[x_{\rm F}y_{\rm F}\Big(x_{\rm F}^{2}+y_{\rm F}^{2}\Big)
       -\ln(x_{\rm F}+y_{\rm F})\right]
       -C_{\ell} \frac{Z^{2}}{A^{4/3}}p^{}_{\rm F} \;.
 \label{EoverA2}
\end{equation}
Note that given $A$, $Z$, and $n\!=\!A_{\rm total}/V$, the only unknown 
quantity in the above expression is the nuclear mass $M(N,Z)$. While
experimentally available for a large number of nuclei around the line
of stability, nuclear masses near the drip line are unknown, thereby
making the need for theoretical extrapolations unavoidable. As crustal
properties become better determined, nuclear masses at the drip line
will be strongly constrained. Alternatively, the advent of facilities
capable of producing beams of rare isotopes to explore the limits of
nuclear existence will place strong constraints on crustal properties.

Having computed the energy per baryon of the system, we are now in a
position to compute two additional thermodynamic properties that are
essential for the understanding of both the structure and composition
of the outer crust. These are the equation of state, namely, the
relation between pressure and density, and the chemical
potential. Recall that in modeling the outer crust the central
assumption is that of thermal, hydrostatic, and chemical equilibrium.
Thus, complete equilibrium demands the equality of temperature,
pressure, and chemical potential at each layer of the outer crust.

At zero temperature and for a constant number of particles, the
pressure of the system may be computed from the total energy of the
system. As the individual nuclei do not contribute to the pressure,
one must only compute the electronic and lattice contributions. In
particular, the electronic contribution at zero temperature is given
by
\begin{equation}
 P_{e}\mathop{=}_{T=0}
     -\left(\frac{\partial E_{e}}{\partial V}\right)_{\!\!Z}
     =\frac{x_{\rm F}}{3}\left(\frac{\partial\mathscr{E}_{e}}
      {\partial x_{\rm F}}\right)-\mathscr{E}_{e} =
      \frac{m_{e}^{4}}{3\pi^{2}}\left(x_{\rm F}^{3}y_{\rm F} -
      \frac{3}{8}
      \left[x_{\rm F}y_{\rm F}\Big(x_{\rm F}^{2}+y_{\rm F}^{2}\Big)
      -\ln(x_{\rm F}+y_{\rm F})\right]\right)\;.
 \label{PElec}
\end{equation}
Similarly, the lattice contribution to the pressure is given by
the following simple expression:
\begin{equation}
 P_{\ell}\mathop{=}_{T=0}
     -\left(\frac{\partial E_{\ell}}{\partial V}\right)_{\!\!A,Z}
   = -\frac{n}{3}C_{\ell} \frac{Z^{2}}{A^{4/3}}p^{}_{\rm F}\;.
 \label{PLatt}
\end{equation}
In this manner the full ({\sl electronic} plus {\sl lattice}) 
contribution to the pressure may be written as
\begin{equation}
 P(A,Z;n)= \frac{m_{e}^{4}}{3\pi^{2}}\left(x_{\rm F}^{3}y_{\rm F} 
            - \frac{3}{8}\left[x_{\rm F}y_{\rm F}\Big(x_{\rm F}^{2}+
              y_{\rm F}^{2}\Big)-\ln(x_{\rm F}+y_{\rm F})\right]\right)
            -\frac{n}{3}C_{\ell} \frac{Z^{2}}{A^{4/3}}p^{}_{\rm F}\;.
 \label{PTotal}
\end{equation}

As alluded earlier, full equilibrium in the system is established by
demanding that the temperature, pressure, and chemical potential ---
but not necessarily the baryon density --- be continuous throughout
the outer crust. As the temperature of the system is assumed to be
equal to zero, the only remaining thermodynamic observable to
calculate is the chemical potential. At zero temperature, the Gibbs
free energy and the total energy of the system are related by a
Legendre transform ($G\!=\!E\!+\!PV$). That is,
\begin{equation}
 \mu(A,Z;P)=\frac{G(A,Z;P)}{A_{\rm total}}  
           =\varepsilon(A,Z;n)+\frac{P}{n}
           =\frac{M(N,Z)}{A}+\frac{Z}{A}\mu_{e}
           -\frac{4}{3}C_{\ell}\frac{Z^{2}}{A^{4/3}}p^{}_{\rm F}\;,
 \label{ChemPotential}
\end{equation}
where $\mu_{e}\!=\!\sqrt{p_{{\rm F}e}^{2}+m_{e}^{2}}$ is the
electronic chemical potential. Note that the chemical potential is a
function of the pressure whereas the energy per baryon is a function
of the baryon density. The transformation from one into the other is
accomplished through Eq.~(\ref{PTotal}). Also note that as hydrostatic
and chemical equilibrium must be maintained throughout the star, the
composition of the outer crust must be obtained by minimizing the
Gibbs free energy per particle ({\it i.e.,} $\mu$) at constant
pressure rather than by minimizing the energy per particle at constant 
baryon density. This procedure will be carried out in the next section.

\section{Results}
\label{Results}
In this section results will be presented for the structure and
composition of the outer crust. The implementation of the ideas
developed in the previous section will be carried out by using various
models for nuclear masses. Two of these models are based on
sophisticated microscopic/macroscopic models that yield
root-mean-square (RMS) errors of only a fraction of an MeV when
compared to large databases of available experimental nuclear
masses~\cite{Audi:1993zb,Audi:1995dz}. These two models are the ones
by Duflo and Zuker~\cite{Duflo:1994,Zuker:1994, Duflo:1995} and the
finite range droplet model of M\"oller, Nix, and
collaborators~\cite{Moller:1993ed,Moller:1997bz}.  The other two
models are based on accurately-calibrated microscopic approaches that
employ a handful of empirical parameters to reproduce the ground-state
properties of finite nuclei and some collective
excitations~\cite{Lalazissis:1996rd,Lalazissis:1999,Todd-Rutel:2005fa}.
While successful, the RMS errors of these two microscopic approaches
are significantly larger than those obtained with the
microscopic/macroscopic models. Yet one of the great advantages of the
microscopic models is the ability to systematically study the impact
of unknown physics on crustal properties. First and foremost, we are
interested in understanding how models that are equally successful in
describing available ground-state properties of finite nuclei differ
in their predictions of exotic (neutron-rich) nuclei.

\subsection{Toy Model of the Outer Crust}
\label{ToyModel}

Although the structure and composition of the crust will be ultimately
computed using sophisticated mass formulas, we start by introducing a
{\sl ``toy model''} that while simple, captures the essential physics 
of the outer crust, namely, a competition between an electronic density 
that drives the system towards more neutron-rich nuclei and a nuclear 
symmetry energy that opposes such a change.

The toy model consists of the following two approximations. First, a
simple liquid-drop model will be used to compute nuclear masses [see
Eq.~(\ref{EoverANucl})]. Second, the electronic contribution will be
assumed to be that of an extremely relativistic ({\it i.e.,}
$m_{e}/p_{{\rm F}e}\!\rightarrow\!0$) Fermi gas. While both of these
approximations will be relaxed in the next section, we believe that
the physical insights that one develops from this analytic treatment
are valuable.

The liquid-drop mass formula may be written in the absence of
pairing correlations (and assuming $Z(Z\!-\!1)\!\approx\!Z^{2}$)
as follows:
\begin{equation}
 \varepsilon_{n}(x,y) = 
  m_{p}y + m_{n}(1-y) - a_{\rm v} + \frac{a_{\rm s}}{x} +
  a_{\rm c}x^{2}y^{2} + a_{\rm a}(1-2y)^{2} \;,
 \label{LiquidDrop}
\end{equation}
where $x\!\equiv\!A^{1/3}$ and $y\!\equiv\!Z/A$ is the proton (or
electron) fraction. The various empirical constants ($a_{\rm v}$, 
$a_{\rm s}$, $a_{\rm c}$, and $a_{\rm a}$) represent volume, surface, 
Coulomb, and asymmetry contribution, respectively. Using a 
least-square fit to 2049 nuclei (available online at the UNEDF 
collaboration website {\tt http://www.unedf.org/}) one obtains the 
following values for the four empirical constants:
\begin{equation}
  a_{\rm v}=15.71511~{\rm MeV}, \;
  a_{\rm s}=17.53638~{\rm MeV}, \;
  a_{\rm c}= 0.71363~{\rm MeV}, \;
  a_{\rm a}=23.37837~{\rm MeV}.
 \label{LiquidDropParams}
\end{equation}

To understand the competition among the various terms --- and to
anticipate how this competition will be modified in the presence
of a Fermi gas of electrons --- we compute the optimal values of
$x$ and $y$ using the simple liquid-drop formula by setting both
derivatives equal to zero. That is,
\begin{subequations}
 \begin{align}
   & \left(\frac{\partial\varepsilon_{n}}{\partial x}\right)_{\!\!\!y} = 
          -\frac{a_{\rm s}}{x^{2}}+2a_{\rm c}xy^{2}=0\;,
           \label{dEpsilonx}\\ 
   & \left(\frac{\partial\varepsilon_{n}}{\partial y}\right)_{\!\!\!x} = 
          -\Delta m+2a_{\rm c}x^{2}y-4a_{\rm a}(1-2y)=0\;,
           \label{dEpsilony}
 \end{align}
 \label{dEpsilon}
\end{subequations}
where we have defined $\Delta m\!\equiv\!m_{n}\!-\!m_{p}$. The above 
set of equations has the following simple analytic solution:
\begin{subequations}
 \begin{align}
   & A=x^{3}=\left(\frac{a_{\rm s}}{2a_{\rm c}}\right)\frac{1}{y^{2}}\;, 
     \label{dEpsilonxSol}\\ 
   & y=\frac{1+\displaystyle{\left(\frac{\Delta m}{4a_{\rm a}}\right)}}
       {2+\displaystyle{\left(\frac{a_{\rm c}}{2a_{\rm a}}\right)x^{2}}} 
       \approx \frac{1/2}{1+\displaystyle{\left(\frac{a_{\rm c}}
       {4a_{\rm a}}\right)x^{2}}}\;. 
     \label{dEpsilonySol}
 \end{align}
 \label{dEpsilonSol}
\end{subequations}
The above solutions suggest the following physical interpretation. 
For a fixed proton fraction $y\!=\!Z/A$, the optimal value of $x$
emerges from a competition between surface (which favors large $x$)
and Coulomb contributions (which favors small $x$). For the set of
empirical constants given in Eq.~(\ref{LiquidDropParams}), the
relevant ratio is given by $a_{\rm s}/2a_{\rm c}\!\approx\!12.287$.
Conversely, if $A\!=\!x^{3}$ is held fixed, then the optimal proton
fraction $y$ results from the competition between Coulomb and
asymmetry contribution, with the former favoring $y\!=\!0$ and
the latter $y\!=\!1/2$. If both equation are solved 
{\sl simultaneously}, then one finds the most stable nucleus for
this parameter set. One obtains $x_{0}\!=\!3.906$ and 
$y_{0}\!=\!0.454$, or equivalently:
\begin{equation}
  A_{0}=59.598,\;  Z_{0}=27.060,\; 
  N_{0}=32.538,\;  (B/A)_{0}=8.784~{\rm MeV},\;
  m_{0}=930.195~{\rm MeV},
 \label{MostStableNucleus}
\end{equation}
with $m_{0}\!\equiv\!(M/A)_{0}$ being the nuclear mass per nucleon.

The second assumption defining the toy model is that of an extremely
relativistic Fermi gas of electrons ({\it i.e.,} $p_{{\rm
F}e}\!\gg\!m_{e}$). In this limit one obtains simple expressions for
the total energy per baryon, chemical potential, and pressure in terms
of the adopted set of variables. That is,
\begin{subequations}
 \begin{align}
  & \varepsilon(x,y,p_{\rm F}^{}) = \varepsilon_{n}(x,y) +
    \frac{3}{4}y^{4/3}p_{\rm F}^{}-C_{\ell}\,x^{2}y^{2}p^{}_{\rm F}\;,
  \label{Energy} \\
  & \mu(x,y,p_{\rm F}^{}) = \varepsilon_{n}(x,y) + y^{4/3}p^{}_{\rm F}
           -\frac{4}{3}C_{\ell}\,x^{2}y^{2}p^{}_{\rm F}\;,
    \label{ChemPot}\\
  & P(x,y,p_{\rm F}^{}) = \frac{n}{4}y^{4/3}p^{}_{\rm F}
  -\frac{n}{3}C_{\ell}\,x^{2}y^{2}p^{}_{\rm F}\;.
    \label{Pressure}
 \end{align}
 \label{EPandMu}
\end{subequations}

Before assessing the quantitative impact of the density dependent
({\it i.e.,} electronic and lattice) contributions on the
semi-empirical mass formula, a few comments are in order. First, at
the predicted neutron-drip density of about $4\!\times\!10^{11}{\rm
g/cm}^{3}$, the Fermi momentum is approximately equal to $p_{\rm
F}^{\rm max}\!\approx\!40$~MeV. This suggests a large electronic
contribution at those densities of about $\varepsilon_{e}^{\rm
max}\!\approx\!30\,y^{4/3}$~MeV. As the nuclear contribution is
independent of density, the electrons will drive the system to small
values of $y$. Second, the lattice contribution (perhaps not
surprisingly) has the same dependence on $x$ and $y$ as the Coulomb
contribution to the semi-empirical mass formula. Indeed, the full
impact of the lattice contribution can be included through a
redefinition, albeit a density dependent one, of the Coulomb
coefficient.  That is, $a_{\rm c}\!\rightarrow\!\widetilde{a}_{c}
(p^{}_{\rm F})\!  \equiv\!(a_{\rm c}-C_{\ell}p^{}_{\rm F})$.  As the 
optimal value of the proton fraction $y$ (for fixed $x$) emerges 
from a competition between Coulomb and asymmetry terms [see 
Eq.~(\ref{dEpsilonySol})], the lattice contribution drives the 
system towards the symmetric $y\!=\!1/2$ limit. Yet the lattice 
contribution is marginal. Indeed, even at neutron-drip densities 
its contribution provides a meager, although by no means negligible,
correction to the dominant Coulomb term. The above two facts summarize
the main structure of the outer crust, namely, a nuclear lattice
embedded in an electron gas that is responsible for driving the system
towards progressively more neutron-rich nuclei. Thus the outer crust
represents a unique laboratory for the study of neutron-rich nuclei in
the $Z\!\approx\!20\!-\!50$ region. As such, it nicely complements
{\sl Rare-Isotope Facilities} worldwide that aim to provide a detailed
map of the nuclear landscape.

Incorporating electronic and lattice contributions to the 
semi-empirical mass formula yields the following expression 
for the total energy per nucleon of the system:
\begin{equation}
 \varepsilon(x,y,p_{\rm F}^{}) =  
  m_{p}y + m_{n}(1-y) - a_{\rm v} + \frac{a_{\rm s}}{x} +
  a_{\rm c}x^{2}y^{2} + a_{\rm a}(1-2y)^{2} +
  \frac{3}{4}y^{4/3}p_{\rm F}^{} - C_{\ell}\,x^{2}y^{2}p^{}_{\rm F}\;.
  \label{LiquidDrop2}
\end{equation}

As done before for the pure nuclear contribution, the optimal values 
of $x$ and $y$ --- at fixed density ---  may be obtained by setting 
both derivatives equal to zero. That is,

\begin{subequations}
 \begin{align}
   & \left(\frac{\partial\varepsilon}{\partial x}
           \right)_{\!\!\!y,p^{}_{\rm F}} = 
          -\frac{a_{\rm s}}{x^{2}}+
            2\widetilde{a}_{\rm c}xy^{2}=0\;,
           \label{dEpsilonx2}\\ 
   & \left(\frac{\partial\varepsilon_{n}}{\partial y}
           \right)_{\!\!\!x,p^{}_{\rm F}} = 
          -\Delta m+2\widetilde{a}_{\rm c}x^{2}y - 4a_{\rm a}(1-2y)
          +y^{1/3}p_{\rm F}^{}=0\;,
           \label{dEpsilony2}
 \end{align}
 \label{dEpsilon2}
\end{subequations}
where a ``renormalized'' Coulomb coefficient has been defined as
\begin{equation}
 \widetilde{a}_{c}(p^{}_{\rm F})\equiv(a_{\rm c}-C_{\ell}p^{}_{\rm F})\;.
  \label{Coulomb2}
\end{equation}

\subsubsection{First-order Solution}
\label{FirstOrder}

Before providing exact solutions to Eqs.~(\ref{dEpsilon2}), we compute
approximate solutions that are accurate to first order in $p_{\rm
F}^{}$.  In addition of being analytic, these closed-form expressions
provide valuable insights into the composition of the outer crust. The
first-order solutions are obtained by incorporating the density
dependence in the following form:
\begin{equation}
  x(p^{}_{\rm F})=x_{0}(1+\xi)  \;\;{\rm and}\;\;
  y(p^{}_{\rm F})=y_{0}(1+\eta) \;,
 \label{FirstOrderExp}
\end{equation}
where both $\xi$ and $\eta$ represent small ({\it i.e.,} first-order
in $p_{\rm F}^{}$) deviations from the zero-density results. Substituting 
the above equations into Eqs.~(\ref{dEpsilon2}) yields the first-order 
solutions. One obtains,
\begin{subequations}
 \begin{align}
 & x(p^{}_{\rm F})=x_{0}\left[1+
  \left(\frac{(C_{1}-1)C_{\ell}+2C_{2}}{3C_{1}-1}\right)
  \frac{p_{\rm F}}{a_{\rm c}}\right] =
  (3.90610+0.03023p^{}_{\rm F}) \;,
  \label{FirstOrderSolsx}\\ 
 & y(p^{}_{\rm F})=y_{0}\left[1-
  \left(\frac{3C_{2}-C_{\ell}}{3C_{1}-1}\right)
   \frac{p_{\rm F}}{a_{\rm c}}\right] =
  (0.45405-0.00419p^{}_{\rm F})\;.
  \label{FirstOrderSolsy}
 \end{align}
 \label{FirstOrderSols}
\end{subequations}
Note that in the above expressions the Fermi momentum should be
given in MeV. Moreover, for simplicity the following two dimensionless 
quantities were introduced:
\begin{equation}
  C_{1} \equiv \frac{4a_{a}}{x_{0}^{2}a_{\rm c}} \approx 8.58843
  \;\;\; {\rm and} \;\;\;
  C_{2} \equiv \frac{1}{2x_{0}^{2}y_{0}^{2/3}} \approx 0.05547 \;.
\end{equation}

The first-order equations [Eqs.~(\ref{FirstOrderSols})] --- while not
necessarily quantitatively accurate --- provide useful insights into
how the composition of the outer crust evolves with density. As
previously suggested, the proton fraction $y$ decreases with density
in an effort to minimize the ``repulsive'' electronic
contribution. Indeed, to an excellent approximation
Eq.~(\ref{FirstOrderSolsy}) may be written in the following simple
form:
\begin{equation}
 y(p^{}_{\rm F})=y_{0}-\frac{p^{}_{{\rm F}e}}{8a_{\rm a}}
              =(0.45405-0.00411p^{}_{\rm F})\;.  
 \label{FirstOrderSolsy2}
\end{equation}
As indicated in Eq.~(\ref{dEpsilonySol}), the optimal value of $y_{0}$
emerges from a competition between Coulomb and asymmetry terms, with
the former driving $y_{0}$ towards zero and the latter towards one half.
The above equation indicates that the evolution of $y$ with density is
controlled by the dimensionless ratio of $p^{}_{{\rm F}e}/a_{a}$,
suggesting that the larger the value of the asymmetry energy, the
slower the evolution away from $y_{0}$; that is, the more symmetric
the nucleus will remain. Moreover, as the denominator
in Eq.~(\ref{FirstOrderSolsy2}) [$8a_{\rm a}\!\approx\!100$~MeV] is
significantly larger than the electronic Fermi momentum over the
entire region of interest, the first-order approximation is expected
to be fairly accurate over the entire outer crust. Indeed, assuming 
a realistic value for the drip density of
$\rho_{\rm drip}\!=\!4\!\times\!10^{11}~{\rm g/cm}^{3}$ yields a
proton fraction of $y^{}_{\rm drip}\!=\!0.298$. This represents a 2\%
deviation from the value of $y({}^{118}{\rm Kr})\!=\!0.305$ for the
conventionally accepted drip nucleus ${}^{118}$Kr.

\begin{figure}[h]
  \includegraphics[width=4in]{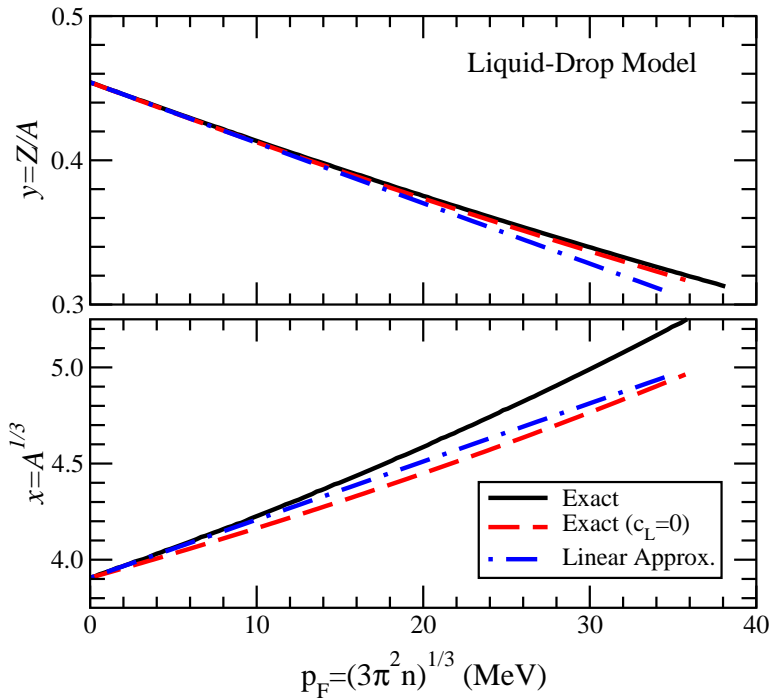}
 \vspace{-0.2cm}
 \caption{(color online)
          Baryon number $x\!=\!A^{1/3}$ (lower panel) and proton
          fraction $y\!=\!Z/A$ (upper panel) are displayed as a 
          function of the Fermi momentum 
          $p^{}_{\rm F}\!\equiv\!(3\pi^{2}n)^{1/3}$. The black 
          solid lines represent the exact solution to the 
          toy-model problem given in Eqs.~(\ref{dEpsilon2}), 
          while the red dashed lines display the corresponding
          solution in the $C_{\ell}\!\equiv\!0$ (no lattice) 
          limit [see Eqs.~(\ref{ToySols})]. Finally, the 
          low-density solution [Eqs.~(\ref{FirstOrderSols})] 
          is displayed by the blue dot-dashed lines.} 
 \label{Fig1}
\end{figure}

\subsubsection{Exact Solution}
\label{ExactToySolution}

The exact solution to the toy-model problem requires (for a fixed
value of $p^{}_{\rm F}$) to find the simultaneous roots of
Eqs.~(\ref{dEpsilonx2}) and~(\ref{dEpsilony2}). While numerically
simple, the exact solution can not be displayed in closed form. Yet
the exact solution differs only slightly from the
$C_{\ell}\!\equiv\!0$ solution --- which has an analytic, albeit a bit
unorthodox, solution. The closed-form solution for the
$C_{\ell}\!\equiv\!0$ case may be obtained by simply re-writing
Eqs.~(\ref{dEpsilon2}). That is,
\begin{subequations}
 \begin{align}
   & x(y)=\left(\frac{a_{\rm s}}{2{a_{\rm c}}y^{2}}\right)^{1/3}\;,
           \label{ToySols1} \\ 
   & p^{}_{\rm F}(y)=\frac{\Delta m-2a_{\rm c}x^{2}y+4a_{\rm a}(1-2y)}
                    {y^{1/3}} \;.  
           \label{ToySols2}
 \end{align}
 \label{ToySols}
\end{subequations}
The above set of equations suggest that rather than looking for a
solution of $x$ and $y$ as a function of $p^{}_{\rm F}$, one should
{\sl ``solve''} for $x$ and $p^{}_{\rm F}$ as a function of $y$, with
the maximum value of $y$ given by $y_{\rm max}\!=\!y_{0}\!=\!0.45405$
and the minimum value of $y$ given by the condition 
$\mu(y_{\rm min})\!=\!m_{n}$. 

In Fig.~\ref{Fig1} the baryon number $x\!=\!A^{1/3}$ and proton
fraction $y\!=\!Z/A$ are displayed as a function of the Fermi 
momentum $p^{}_{\rm F}\!\equiv\!(3\pi^{2}n)^{1/3}$. The black solid 
line display the exact numerical solution to the toy-model problem 
[see Eqs.~(\ref{dEpsilon2})]. In this simple model, the drip line 
density is predicted to be at 
$\rho_{\rm drip}\!=\!4\!\times\!10^{11}~{\rm g/cm}^{3}$ with the 
drip-line nucleus being ${}^{154}$Cd ({\it i.e.,} $Z\!=\!48$ and 
$N\!=\!106$). The solution obtained by ignoring the lattice
contribution is displayed by the red dashed line. Because the
lattice contribution to the chemical potential is negative, the
$C_{\ell}\!\equiv\!0$ solution reaches the drip line faster,
{\it i.e.,} at a lower density. Moreover, as the lattice
contribution {\sl ``renormalizes''} the Coulomb term in the
semi-empirical mass formula (or equivalently, enhances the role 
of the symmetry energy) the $C_{\ell}\!\equiv\!0$ solution predicts 
a lower proton fraction than the exact solution. Finally, the 
dot-dashed blue line displays the solution correct to first-order 
in $p^{}_{\rm F}$. In the particular case of the proton fraction 
$y$, the approximate linear solution 
$y=y_{0}-p^{}_{{\rm F}e}/8a_{\rm a}$ [Eq.~(\ref{FirstOrderSolsy2})]
reproduces fairly accurately the behavior of the exact solution.

\begin{figure}[h]
  \includegraphics[width=4in]{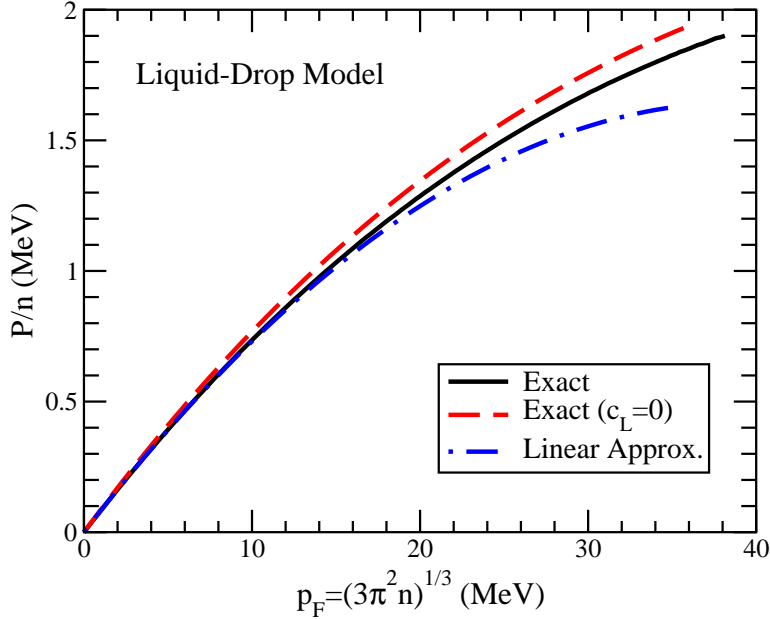}
 \vspace{-0.2cm}
 \caption{(color online)
          Pressure as a function of the Fermi momentum 
          $p^{}_{\rm F}\!\equiv\!(3\pi^{2}n)^{1/3}$. The black 
          solid line represent the exact solution to the 
          toy-model problem, the red dashed line displays 
          the corresponding $C_{\ell}\!\equiv\!0$ (no lattice)
          solution, and the low-density solution is displayed 
          by the blue dot-dashed line.}
 \label{Fig2}
\end{figure}

The equation of state ({\it i.e.,} pressure {\it vs} density)  
predicted by the toy model is displayed in Fig.~\ref{Fig2}. 
As the lattice provides a negative contribution to the pressure 
[Eq.~(\ref{Pressure})], the equation of state for the 
$C_{\ell}\!\equiv\!0$ case is slightly stiffer than the 
exact one. The first-order solution in $p^{}_{\rm F}$ provides 
a {\sl quantitatively accurate} description of the equation of 
state up to fairly large values of the density. Note that the 
first-order approximation to the pressure is defined as follows:
\begin{equation}
 \frac{P}{np^{}_{\rm F}}=\frac{1}{4}y^{4/3}
           -\frac{1}{3}C_{\ell}\,x^{2}y^{2}
 \approx (0.08367-0.00106p^{}_{\rm F})\;.  
 \label{FirstOrderP}
\end{equation}

\subsection{Realistic Models of the Outer Crust}
\label{RealisticModel}
 
In this section we employ realistic nuclear mass models to compute the
structure and composition of the outer crust. Two of the
models~\cite{Moller:1993ed,Moller:1997bz,Duflo:1994,Zuker:1994,Duflo:1995}
are based on sophisticated mass formulas that have been calibrated to
thousands of available experimental masses throughout the periodic
table~\cite{Audi:1993zb,Audi:1995dz}. The other two models are based 
on accurately-calibrated microscopic approaches that employ a handful 
of empirical parameters to reproduce the ground-state properties of 
finite nuclei and some nuclear collective 
excitations~\cite{Lalazissis:1996rd,Lalazissis:1999,Todd-Rutel:2005fa}.

\begin{figure}[h]
  \includegraphics[width=4in]{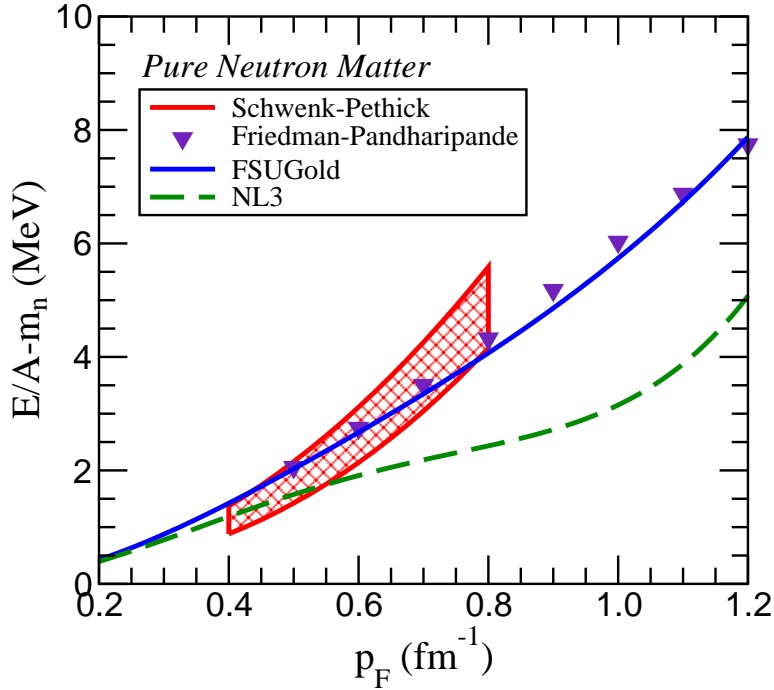}
 \vspace{-0.2cm}
 \caption{(color online) Energy per particle for pure neutron
         as a function of the neutron Fermi momentum. Shown are 
         the microscopic model of Friedman and Pandharipande
         ~\cite{Friedman:1981qw} (purple triangles) and the 
         {\sl model-independent} result based on the physics 
         of resonant Fermi gases by Schwenk and Pethick
         ~\cite{Schwenk:2005ka} (red region). Also shown are
         the predictions from the accurately calibrated
         NL3~\cite{Lalazissis:1996rd,Lalazissis:1999} (green 
         dashed line) and  FSUGold~\cite{Todd-Rutel:2005fa} 
         (blue line) parameter sets.} 
 \label{Fig3}
\end{figure}

Whereas the microscopic models have yet to attained the level of
precision displayed by the microscopic/macroscopic ones, they are
valuable in elucidating various details of the underlying physics.  
For example, in the previous section we established the critical 
role played by the symmetry energy in the evolution of the proton 
fraction with density [see Eq.~(\ref{FirstOrderSolsy2})]. However, 
it is unknown how the symmetry energy coefficient $a_{\rm a}$ 
changes as nuclei move towards the drip line. Presumably, the 
development of a significant neutron skin makes these nuclei (on 
average) more dilute than their stable counterparts. If so, one 
needs to extrapolate the symmetry energy to lower densities, a 
procedure that is highly uncertain because of our poor knowledge 
of the \emph{slope} of the symmetry energy. To illustrate this
uncertainty, the equation of state of pure neutron matter predicted 
by NL3 (green dashed line) and FSUGold (blue solid line) is displayed 
in Fig.~\ref{Fig3}. For comparison, we also show the predictions
from the microscopic model of Friedman and Pandharipande based on 
realistic two-body interactions~\cite{Friedman:1981qw} (purple 
upside-down triangles) and the \emph{model-independent} result 
based on the physics of resonant Fermi gases by Schwenk and 
Pethick~\cite{Schwenk:2005ka} (red hatched region). Note that to a
very good approximation, the equation of state of pure neutron matter
equals that of symmetric nuclear matter \emph{plus} the symmetry
energy. The differences between NL3 and FSUGold displayed in
Fig.~\ref{Fig3} are {\sl all} due to the large uncertainties
in the symmetry energy. In particular, as NL3 predicts a stiffer 
equation of state than FSUGold, namely, one whose energy increases 
faster with density {\sl at high densities}, the symmetry energy of 
NL3 is lower than that of FSUGold at sub-saturation densities. Thus, 
FSUGold has been shown to reach the neutron-drip lines earlier than
NL3~\cite{Todd:2003xs}. By the same token, NL3 should predict a
sequence of more neutron-rich nuclei (lower $y$) in the outer
crust than FSUGold.

\begin{figure}[h]
  \includegraphics[width=4.5in]{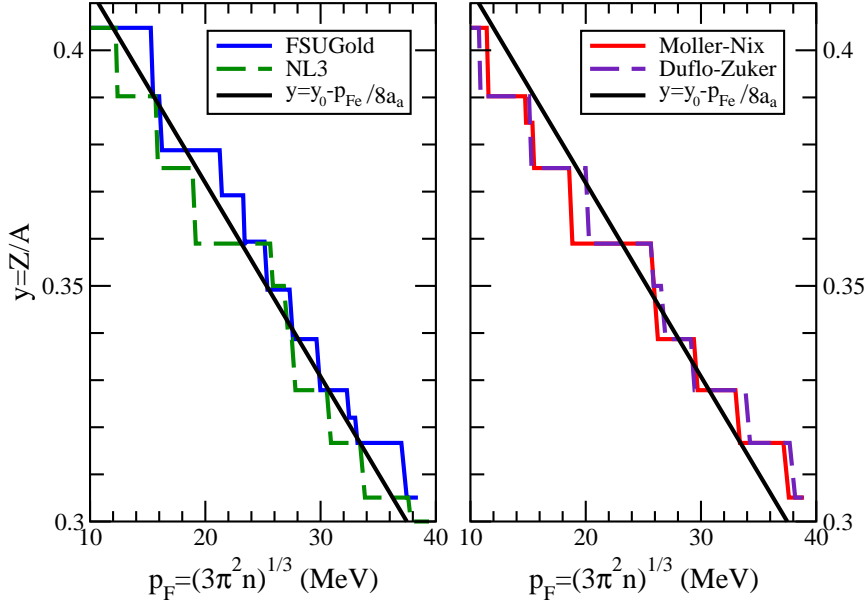}
 \vspace{-0.2cm}
 \caption{(color online) Shown in the left-hand panel is the proton 
          fraction predicted by the accurately calibrated FSUGold 
          (blue solid line) and NL3 (green dashed line) parameter 
          sets. Also shown is the simple liquid-drop formula given 
          in Eq.~(\ref{FirstOrderSolsy2}). The right-hand panel 
          displays the corresponding proton fraction as predicted 
          by the mass formulas from Moller-Nix (red solid line) 
          and Duflo-Zuker (purple dashed line).} 
 \label{Fig4}
\end{figure}

Shown in the left-hand panel of Fig.~\ref{Fig4} is the proton fraction
predicted by the two microscopic models; FSUGold (blue solid line) and
NL3 (green dashed line). Also shown is the simple prediction obtained
from the liquid-drop formula [Eq.~(\ref{FirstOrderSolsy2})]. The
proton fraction predicted with the FSUGold parameter set is
consistently higher than for the NL3 set. This is a reflection of the
stiffer penalty imposed on the FSUGold set for departing from the
symmetric ($N\!=\!Z$) limit. The right-hand panel shows the 
corresponding behavior for the case of the microscopic/macroscopic
models of Moller-Nix (red solid line) and Duflo-Zuker (purple dashed
line). Differences among these two models are small.

\begin{figure}[h]
  \includegraphics[width=4in]{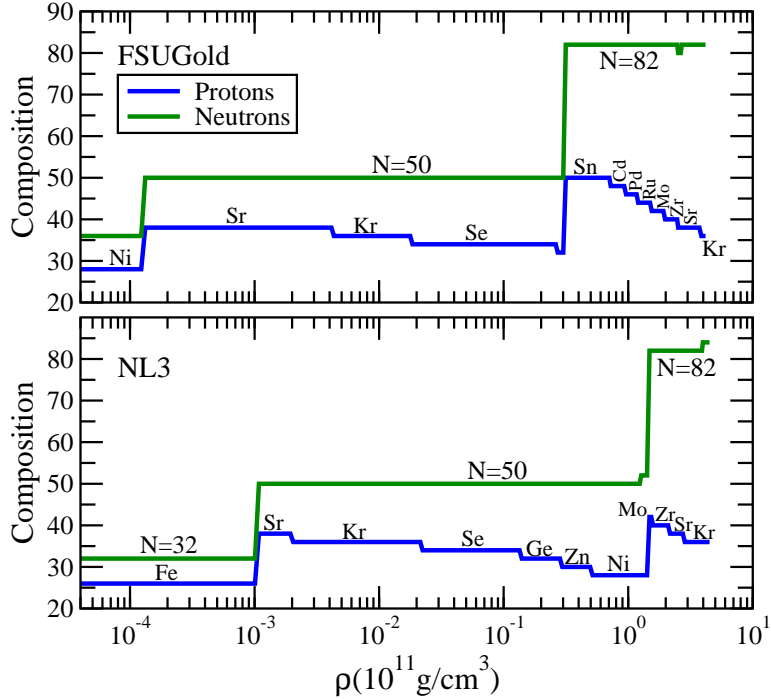}
 \vspace{-0.2cm}
 \caption{(color online) Composition of the outer crust of a 
           neutron star as predicted by the accurately calibrated 
           FSUGold (upper panel) and NL3 (lower panel) parameter 
           sets. Protons are displayed with the (lower) blue line 
           while neutrons with the (upper) green line.}
 \label{Fig5}
\end{figure}

\begin{figure}[h]
  \includegraphics[width=4in]{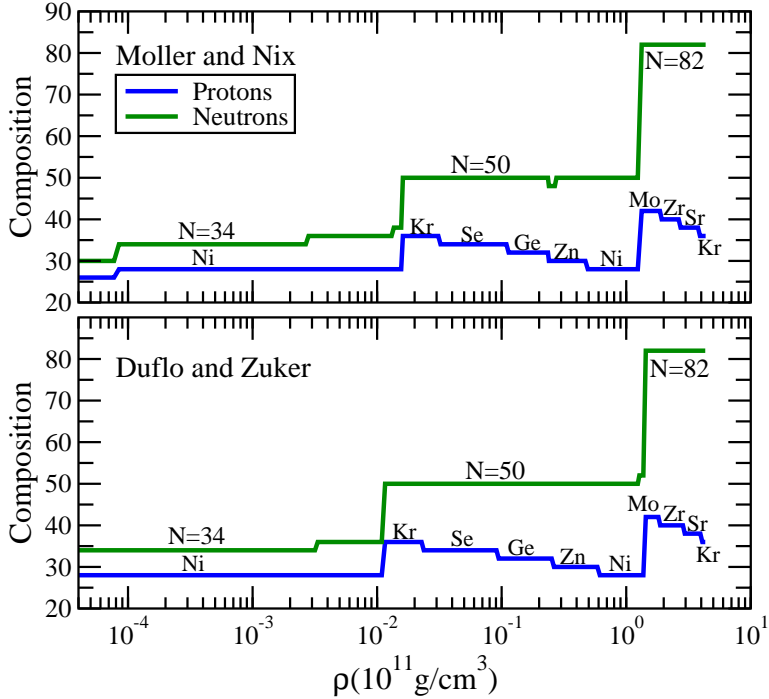}
 \vspace{-0.2cm}
 \caption{(color online) Composition of the outer crust of a 
           neutron star as predicted using the the mass formulas
           of Moller-Nix (upper panel) and Duflo-Zuker (lower panel). 
           Protons are displayed with the blue (lower) line 
           while neutrons with the green (upper) line.}
 \label{Fig6}
\end{figure}

Similar trends may be observed in Figs.~\ref{Fig5} and~\ref{Fig6}
where the composition of the outer crust is displayed as a function of
density. As the system makes a rapid jump in neutron number (say to
magic number $N\!=\!50$) the proton number jumps with it. Along this
neutron plateau, the proton fraction decreases systematically with
increasing density in an effort to reduce the electronic contribution
to the chemical potential.  Eventually, the neutron-proton mismatch is
too large and the jump to the next neutron plateau ensues; a jump that
is driven by the symmetry energy. Clearly, the larger the symmetry
energy at low densities, the smaller the neutron-proton mismatch and
the early the jump to the next neutron plateau. These features are
clearly displayed in Fig.~\ref{Fig5} as one contrasts the behavior of
FSUGold to that of NL3. In contrast, few differences are noticeable 
in Fig.~\ref{Fig6} when comparing the model of Moller-Nix to that
of Duflo-Zuker.

\begin{figure}[h]
  \includegraphics[width=4in]{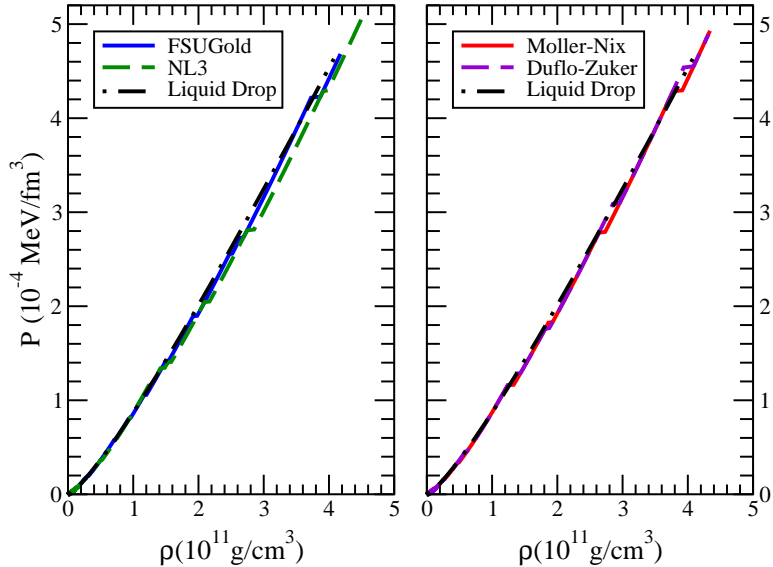}
 \vspace{-0.2cm}
 \caption{(color online) Shown in the left-hand panel is the 
          zero-temperature equation of state (pressure {\it vs}
          density) predicted by the accurately calibrated FSUGold 
          (blue solid line) and NL3 (green dashed line) parameter 
          sets. Also shown is the prediction from the simple 
          liquid-drop formula. The right-hand panel displays the 
          corresponding expression as predicted by the mass formulas 
          from Moller-Nix (red solid line) and Duflo-Zuker (purple 
          dashed line).} 
 \label{Fig7}
\end{figure}

We conclude this section by displaying in Fig.~\ref{Fig7} 
equation-of-state (pressure {\it vs} density) predictions for the
outer crust of a neutron star. The left-hand panel shows results from
calculations using the FSUGold (blue solid line) and NL3 (green dashed
line) parameter sets. Although barely visible, the density shows
discontinuities at those places where the composition changes
abruptly. It is also noted that the FSUGold parametrization predicts a
pressure that rises slightly faster with density than NL3. For the NL3
set, the symmetry energy admits lower values of the proton/electron
fraction $y$ which, in turn, lowers the pressure of the system. Lower
values of $y$ also yield lower values of the chemical potential,
thereby delaying the arrival to the neutron-drip line. Indeed, 
whereas FSUGold predicts a drip-line density of 
$\rho_{\rm drip}\!=\!4.17\!\times\!10^{11}{\rm g/cm}^{3}$, with
NL3 the transition is delayed by about 8\%, or until
$\rho_{\rm drip}\!=\!4.49\!\times\!10^{11}{\rm g/cm}^{3}$.
A similar plot is shown for the microscopic/macroscopic models of
Moller-Nix (red solid line) and Duflo-Zuker (purple dashed line).
Differences among these two models are barely noticeable. Indeed,
drip-line densities in both models are predicted at about $\rho_{\rm
drip}\!=\!4.3\!\times\!10^{11}{\rm g/cm}^{3}$. Model predictions for
various observables at the base of the outer crust ({\it i.e.,} in the
drip-line region) are listed in Table~\ref{Table1}.

\begin{table}
\begin{tabular}{|c||c|c|c|c|c|c|}
 \hline
 Model & $\rho(10^{11}{\rm g/cm}^{3})$ & $ n(10^{-4}{\rm fm}^{-3})$ 
       & $P(10^{-4}{\rm MeV/fm}^{3})$ & $\mu_{e}({\rm MeV})$  & Element 
       & $B/A({\rm MeV})$ \\ 
 \hline
 \hline
 Moller-Nix  & 4.34 & 2.60 & 4.93 & 26.22 & $^{118}$Kr & $7.21$ \\
 Duflo-Zuker & 4.32 & 2.58 & 4.89 & 26.17 & $^{118}$Kr & $7.19$ \\
 FSUGold     & 4.17 & 2.50 & 4.68 & 25.88 & $^{118}$Kr & $7.11$ \\
 NL3         & 4.49 & 2.69 & 5.06 & 26.39 & $^{120}$Kr & $7.13$ \\
\hline
\end{tabular}
\caption{Equation-of-state observables (mass density, baryon density, 
         pressure, and electronic chemical potential) and composition
         (nucleus and binding-energy per nucleon) at the base of the 
         outer crust.}
\label{Table1}
\end{table}

\section{Conclusions}
\label{Conclusions}

Following the seminal work by Baym, Pethick, and Sutherland, as 
well as the more recent comprehensive work by Ruester, Hempel, and
Schaffner-Bielich, we studied the composition and equation of state of
the outer crust of non-accreting neutron stars. The central focus of
our study was the sensitivity of crustal properties to the density
dependence of the symmetry energy. To do so, four different models
were adopted.  Two of these models, Moller-Nix and Duflo-Zuker, are
based on a combined microscopic/macroscopic approach and yield the
most accurate nuclear masses available in the literature. The other
two models, NL3 and FSUGold, are of a purely microscopic nature and
based on a relativistic mean-field approach. Although the former are
significantly more accurate than the latter, microscopic models have
the advantage of making definite predictions on how the symmetry
energy changes with density (see Fig.~\ref{Fig3}). One can then study
the impact of various features of the symmetry energy --- such as its
slope --- on crustal properties.

The composition and equation of state of the outer crust emerge from a
competition among three relatively simple contributions to the total
energy (or chemical potential) of the system: nuclear, electronic, and
lattice.  The nuclear contribution appears exclusively in the form of
nuclear masses and is independent of the baryon density. The electronic 
contribution is modeled after a zero-temperature free Fermi gas and 
dominates the behavior of the system with baryon density. Finally, the
(body-centered-cubic) lattice contribution is also density dependent
and provides a relatively modest correction (of no more than 10\%) to
the energy of the system. The alluded competition is then primarily
driven by an electronic term that favors a small electron fraction (to
reduce the electronic Fermi energy) and a nuclear symmetry energy that
opposes such a shift towards progressively more neutron-rich nuclei. To
motivate the discussion and to highlight this competition, we
implemented a ``toy model'' of the outer crust by using a simple
semi-empirical ({\sl ``Bethe-Weizs\"acker''}) nuclear mass
formula. Volume, surface, Coulomb, and asymmetry terms were extracted
from a least-square fit to 2049 nuclei (see {\tt
http://www.unedf.org/}). The advantage of such a simple model is that
useful insights emerge from the analytic structure of our
results. Indeed, a particularly transparent result that illustrates
nicely the competition between the electronic contribution and the
nuclear symmetry energy was obtained, namely,
\begin{equation}
 y(p^{}_{\rm F})=y_{0}-\frac{p^{}_{{\rm F}e}}{8a_{\rm a}}
               +{\mathcal O}(p^{2}_{{\rm F}e})\;,
\end{equation}
where $y_{0}$ is the zero-density proton fraction, $p^{}_{{\rm F}e}$ is the 
electronic Fermi momentum and $a_{\rm a}$ is the symmetry energy coefficient. 
While illuminating, this (first-order) result is also surprisingly accurate, 
as the electronic Fermi momentum at the base of the outer crust is very 
close in value to the symmetry energy coefficient 
($p^{}_{{\rm F}e}\!\approx\!26$~MeV {\it vs} $a_{\rm a}\!\approx\!23$~MeV). 
In particular, the toy model predicts a value for the electron fraction 
at the base of the crust that differs by only a few percent from that of
the drip-line nucleus ${}^{118}$Kr.

What is unknown, however, is how the symmetry energy coefficient
$a_{\rm a}$ is modified as nuclei move away from the line of
stability.  Presumably, the symmetry energy is reduced in
neutron-drip nuclei due to the development of a dilute neutron
skin. To investigate the sensitivity of the outer crust to the density
dependence of the symmetry energy we employed two relativistic
mean-field models (NL3 and FSUGold) that while accurately calibrated,
predict a significantly different density dependence for the symmetry
energy. In particular, NL3 predicts a smaller symmetry energy than
FSUGold at the (small) densities of relevance to the outer crust (see
Fig.~\ref{Fig3}). One of the main goals of the present manuscript was
to document how such differences impact the composition of the outer
crust.  One noticed, quite generally, that as the density increases
along a fixed neutron-number plateau (say at magic number $N\!=\!50$) the
proton fraction decreased systematically in an effort to reduce the
electronic contribution to the chemical potential. Eventually,
however, the proton fraction becomes too low and the symmetry energy
drives the system into the next plateau (say at magic number
$N\!=\!82$). How low can the proton fraction get is then 
a question that must be answered by the symmetry energy. Indeed, 
whereas NL3 predicts the formation of ${}^{78}_{28}$Ni${}^{}_{50}$, FSUGold
(having a larger symmetry energy) leaves the $N\!=\!50$ plateau with
the formation of ${}^{82}_{32}$Ge${}^{}_{50}$ (or four protons
earlier). This result may be stated in the form of a correlation
between the neutron radius of ${}^{208}$Pb and the composition of the
outer crust: {\emph{the larger the neutron skin of ${}^{208}${\rm Pb},
the more exotic the composition of the outer crust}}. Finally, and as
it was done in Ref.~\cite{Ruester:2005fm}, we have computed crustal
properties using two of the most accurate tables of nuclear masses
available today, namely, those of Moller-Nix and Duflo-Zuker. Our
results using the model of Moller and Nix agree well with those
published in Ref.~\cite{Ruester:2005fm}. These results are practically
indistinguishable from the ones obtained using the Duflo-Zuker nuclear
mass table; a table that includes 9210 nuclei!

In summary, we have used realistic nuclear mass tables to elucidate
the role of the symmetry energy on the structure and composition of
the outer crust of neutron stars. Recent observations of crustal modes
in magnetars are likely to provide stringent limits on the equation of
state of neutron-rich matter. As the field of nuclear astrophysics
continues to advance --- with the commission of both radioactive beam
facilities as well as ground- and space-based telescopes --- we enter
a new era that promises great hope in the determination of the nuclear
matter equation of state.

\begin{acknowledgments}
The authors are grateful to Jos\'e Barea, Alejandro Frank, 
Jorge Hirsch, and their students for useful discussions and 
for making the Duflo-Zuker mass formula available to us. We 
are also grateful to George Lalazissis for providing the NL3
mass formula. Xavier Roca-Maza acknowledges support from 
grants AP2005-4751 and FIS2005-03142 from MEC (Spain) and
FEDER. This work was supported in part by the Department of 
Energy grant DE-FD05-92ER40750.
\end{acknowledgments}

\vfill\eject
\bibliography{ReferencesJP}

\begin{thebibliography}{45}
\expandafter\ifx\csname natexlab\endcsname\relax\def\natexlab#1{#1}\fi
\expandafter\ifx\csname bibnamefont\endcsname\relax
  \def\bibnamefont#1{#1}\fi
\expandafter\ifx\csname bibfnamefont\endcsname\relax
  \def\bibfnamefont#1{#1}\fi
\expandafter\ifx\csname citenamefont\endcsname\relax
  \def\citenamefont#1{#1}\fi
\expandafter\ifx\csname url\endcsname\relax
  \def\url#1{\texttt{#1}}\fi
\expandafter\ifx\csname urlprefix\endcsname\relax\def\urlprefix{URL }\fi
\providecommand{\bibinfo}[2]{#2}
\providecommand{\eprint}[2][]{\url{#2}}

\bibitem[{\citenamefont{Lattimer and Prakash}(2001)}]{Lattimer:2000nx}
\bibinfo{author}{\bibfnamefont{J.~M.} \bibnamefont{Lattimer}} \bibnamefont{and}
  \bibinfo{author}{\bibfnamefont{M.}~\bibnamefont{Prakash}},
  \bibinfo{journal}{Astrophys. J.} \textbf{\bibinfo{volume}{550}},
  \bibinfo{pages}{426} (\bibinfo{year}{2001}), \eprint{astro-ph/0002232}.

\bibitem[{\citenamefont{Lattimer and Prakash}(2004)}]{Lattimer:2004pg}
\bibinfo{author}{\bibfnamefont{J.~M.} \bibnamefont{Lattimer}} \bibnamefont{and}
  \bibinfo{author}{\bibfnamefont{M.}~\bibnamefont{Prakash}},
  \bibinfo{journal}{Science} \textbf{\bibinfo{volume}{304}},
  \bibinfo{pages}{536} (\bibinfo{year}{2004}), \eprint{astro-ph/0405262}.

\bibitem[{\citenamefont{Baym et~al.}(1971)\citenamefont{Baym, Pethick, and
  Sutherland}}]{Baym:1971pw}
\bibinfo{author}{\bibfnamefont{G.}~\bibnamefont{Baym}},
  \bibinfo{author}{\bibfnamefont{C.}~\bibnamefont{Pethick}}, \bibnamefont{and}
  \bibinfo{author}{\bibfnamefont{P.}~\bibnamefont{Sutherland}},
  \bibinfo{journal}{Astrophys. J.} \textbf{\bibinfo{volume}{170}},
  \bibinfo{pages}{299} (\bibinfo{year}{1971}).

\bibitem[{\citenamefont{Ravenhall et~al.}(1983)\citenamefont{Ravenhall,
  Pethick, and Wilson}}]{Ravenhall:1983uh}
\bibinfo{author}{\bibfnamefont{D.~G.} \bibnamefont{Ravenhall}},
  \bibinfo{author}{\bibfnamefont{C.~J.} \bibnamefont{Pethick}},
  \bibnamefont{and} \bibinfo{author}{\bibfnamefont{J.~R.}
  \bibnamefont{Wilson}}, \bibinfo{journal}{Phys. Rev. Lett.}
  \textbf{\bibinfo{volume}{50}}, \bibinfo{pages}{2066} (\bibinfo{year}{1983}).

\bibitem[{\citenamefont{Hashimoto et~al.}(1984)\citenamefont{Hashimoto, Seki,
  and Yamada}}]{Hashimoto:1984}
\bibinfo{author}{\bibfnamefont{M.}~\bibnamefont{Hashimoto}},
  \bibinfo{author}{\bibfnamefont{H.}~\bibnamefont{Seki}}, \bibnamefont{and}
  \bibinfo{author}{\bibfnamefont{M.}~\bibnamefont{Yamada}},
  \bibinfo{journal}{Prog. Theor. Phys.} \textbf{\bibinfo{volume}{71}},
  \bibinfo{pages}{320} (\bibinfo{year}{1984}).

\bibitem[{\citenamefont{Alford et~al.}(1999)\citenamefont{Alford, Rajagopal,
  and Wilczek}}]{Alford:1998mk}
\bibinfo{author}{\bibfnamefont{M.~G.} \bibnamefont{Alford}},
  \bibinfo{author}{\bibfnamefont{K.}~\bibnamefont{Rajagopal}},
  \bibnamefont{and} \bibinfo{author}{\bibfnamefont{F.}~\bibnamefont{Wilczek}},
  \bibinfo{journal}{Nucl. Phys.} \textbf{\bibinfo{volume}{B537}},
  \bibinfo{pages}{443} (\bibinfo{year}{1999}), \eprint{hep-ph/9804403}.

\bibitem[{\citenamefont{Rajagopal and Wilczek}(2000)}]{Rajagopal:2000wf}
\bibinfo{author}{\bibfnamefont{K.}~\bibnamefont{Rajagopal}} \bibnamefont{and}
  \bibinfo{author}{\bibfnamefont{F.}~\bibnamefont{Wilczek}}
  (\bibinfo{year}{2000}), \eprint{hep-ph/0011333}.

\bibitem[{\citenamefont{Schwenk and Pethick}(2005)}]{Schwenk:2005ka}
\bibinfo{author}{\bibfnamefont{A.}~\bibnamefont{Schwenk}} \bibnamefont{and}
  \bibinfo{author}{\bibfnamefont{C.~J.} \bibnamefont{Pethick}},
  \bibinfo{journal}{Phys. Rev. Lett.} \textbf{\bibinfo{volume}{95}},
  \bibinfo{pages}{160401} (\bibinfo{year}{2005}), \eprint{nucl-th/0506042}.

\bibitem[{\citenamefont{Piekarewicz}(2007)}]{Piekarewicz:2007dx}
\bibinfo{author}{\bibfnamefont{J.}~\bibnamefont{Piekarewicz}},
  \bibinfo{journal}{Phys. Rev.} \textbf{\bibinfo{volume}{C76}},
  \bibinfo{pages}{064310} (\bibinfo{year}{2007}), \eprint{0709.2699}.

\bibitem[{\citenamefont{Brown}(2000)}]{Brown:2000}
\bibinfo{author}{\bibfnamefont{B.~A.} \bibnamefont{Brown}},
  \bibinfo{journal}{Phys. Rev. Lett.} \textbf{\bibinfo{volume}{85}},
  \bibinfo{pages}{5296} (\bibinfo{year}{2000}).

\bibitem[{\citenamefont{Furnstahl}(2002)}]{Furnstahl:2001un}
\bibinfo{author}{\bibfnamefont{R.~J.} \bibnamefont{Furnstahl}},
  \bibinfo{journal}{Nucl. Phys.} \textbf{\bibinfo{volume}{A706}},
  \bibinfo{pages}{85} (\bibinfo{year}{2002}), \eprint{nucl-th/0112085}.

\bibitem[{\citenamefont{Lalazissis et~al.}(1997)\citenamefont{Lalazissis,
  Konig, and Ring}}]{Lalazissis:1996rd}
\bibinfo{author}{\bibfnamefont{G.~A.} \bibnamefont{Lalazissis}},
  \bibinfo{author}{\bibfnamefont{J.}~\bibnamefont{Konig}}, \bibnamefont{and}
  \bibinfo{author}{\bibfnamefont{P.}~\bibnamefont{Ring}},
  \bibinfo{journal}{Phys. Rev.} \textbf{\bibinfo{volume}{C55}},
  \bibinfo{pages}{540} (\bibinfo{year}{1997}), \eprint{nucl-th/9607039}.

\bibitem[{\citenamefont{Lalazissis et~al.}(1999)\citenamefont{Lalazissis,
  Raman, and Ring}}]{Lalazissis:1999}
\bibinfo{author}{\bibfnamefont{G.~A.} \bibnamefont{Lalazissis}},
  \bibinfo{author}{\bibfnamefont{S.}~\bibnamefont{Raman}}, \bibnamefont{and}
  \bibinfo{author}{\bibfnamefont{P.}~\bibnamefont{Ring}}, \bibinfo{journal}{At.
  Data Nucl. Data Tables} \textbf{\bibinfo{volume}{71}}, \bibinfo{pages}{1}
  (\bibinfo{year}{1999}).

\bibitem[{\citenamefont{Todd-Rutel and Piekarewicz}(2005)}]{Todd-Rutel:2005fa}
\bibinfo{author}{\bibfnamefont{B.~G.} \bibnamefont{Todd-Rutel}}
  \bibnamefont{and}
  \bibinfo{author}{\bibfnamefont{J.}~\bibnamefont{Piekarewicz}},
  \bibinfo{journal}{Phys. Rev. Lett} \textbf{\bibinfo{volume}{95}},
  \bibinfo{pages}{122501} (\bibinfo{year}{2005}), \eprint{nucl-th/0504034}.

\bibitem[{\citenamefont{Horowitz et~al.}(2001)\citenamefont{Horowitz, Pollock,
  Souder, and Michaels}}]{Horowitz:1999fk}
\bibinfo{author}{\bibfnamefont{C.~J.} \bibnamefont{Horowitz}},
  \bibinfo{author}{\bibfnamefont{S.~J.} \bibnamefont{Pollock}},
  \bibinfo{author}{\bibfnamefont{P.~A.} \bibnamefont{Souder}},
  \bibnamefont{and} \bibinfo{author}{\bibfnamefont{R.}~\bibnamefont{Michaels}},
  \bibinfo{journal}{Phys. Rev.} \textbf{\bibinfo{volume}{C63}},
  \bibinfo{pages}{025501} (\bibinfo{year}{2001}), \eprint{nucl-th/9912038}.

\bibitem[{\citenamefont{Michaels et~al.}(2005)\citenamefont{Michaels, Souder,
  and Urciuoli}}]{Michaels:2005}
\bibinfo{author}{\bibfnamefont{R.}~\bibnamefont{Michaels}},
  \bibinfo{author}{\bibfnamefont{P.~A.} \bibnamefont{Souder}},
  \bibnamefont{and} \bibinfo{author}{\bibfnamefont{G.~M.}
  \bibnamefont{Urciuoli}} (\bibinfo{year}{2005}),
  \urlprefix\url{http://hallaweb.jlab.org/parity/prex}.

\bibitem[{\citenamefont{Horowitz and
  Piekarewicz}(2001{\natexlab{a}})}]{Horowitz:2000xj}
\bibinfo{author}{\bibfnamefont{C.~J.} \bibnamefont{Horowitz}} \bibnamefont{and}
  \bibinfo{author}{\bibfnamefont{J.}~\bibnamefont{Piekarewicz}},
  \bibinfo{journal}{Phys. Rev. Lett.} \textbf{\bibinfo{volume}{86}},
  \bibinfo{pages}{5647} (\bibinfo{year}{2001}{\natexlab{a}}),
  \eprint{astro-ph/0010227}.

\bibitem[{\citenamefont{Horowitz and
  Piekarewicz}(2001{\natexlab{b}})}]{Horowitz:2001ya}
\bibinfo{author}{\bibfnamefont{C.~J.} \bibnamefont{Horowitz}} \bibnamefont{and}
  \bibinfo{author}{\bibfnamefont{J.}~\bibnamefont{Piekarewicz}},
  \bibinfo{journal}{Phys. Rev.} \textbf{\bibinfo{volume}{C64}},
  \bibinfo{pages}{062802} (\bibinfo{year}{2001}{\natexlab{b}}),
  \eprint{nucl-th/0108036}.

\bibitem[{\citenamefont{Horowitz and Piekarewicz}(2002)}]{Horowitz:2002mb}
\bibinfo{author}{\bibfnamefont{C.~J.} \bibnamefont{Horowitz}} \bibnamefont{and}
  \bibinfo{author}{\bibfnamefont{J.}~\bibnamefont{Piekarewicz}},
  \bibinfo{journal}{Phys. Rev.} \textbf{\bibinfo{volume}{C66}},
  \bibinfo{pages}{055803} (\bibinfo{year}{2002}), \eprint{nucl-th/0207067}.

\bibitem[{\citenamefont{Carriere et~al.}(2003)\citenamefont{Carriere, Horowitz,
  and Piekarewicz}}]{Carriere:2002bx}
\bibinfo{author}{\bibfnamefont{J.}~\bibnamefont{Carriere}},
  \bibinfo{author}{\bibfnamefont{C.~J.} \bibnamefont{Horowitz}},
  \bibnamefont{and}
  \bibinfo{author}{\bibfnamefont{J.}~\bibnamefont{Piekarewicz}},
  \bibinfo{journal}{Astrophys. J.} \textbf{\bibinfo{volume}{593}},
  \bibinfo{pages}{463} (\bibinfo{year}{2003}), \eprint{nucl-th/0211015}.

\bibitem[{\citenamefont{Steiner et~al.}(2005)\citenamefont{Steiner, Prakash,
  Lattimer, and Ellis}}]{Steiner:2004fi}
\bibinfo{author}{\bibfnamefont{A.~W.} \bibnamefont{Steiner}},
  \bibinfo{author}{\bibfnamefont{M.}~\bibnamefont{Prakash}},
  \bibinfo{author}{\bibfnamefont{J.~M.} \bibnamefont{Lattimer}},
  \bibnamefont{and} \bibinfo{author}{\bibfnamefont{P.~J.} \bibnamefont{Ellis}},
  \bibinfo{journal}{Phys. Rept.} \textbf{\bibinfo{volume}{411}},
  \bibinfo{pages}{325} (\bibinfo{year}{2005}), \eprint{nucl-th/0410066}.

\bibitem[{\citenamefont{Piro}(2005)}]{Piro:2005jf}
\bibinfo{author}{\bibfnamefont{A.~L.} \bibnamefont{Piro}},
  \bibinfo{journal}{Astrophys. J.} \textbf{\bibinfo{volume}{634}},
  \bibinfo{pages}{L153} (\bibinfo{year}{2005}), \eprint{astro-ph/0510578}.

\bibitem[{\citenamefont{Strohmayer and Watts}(2006)}]{Strohmayer:2006py}
\bibinfo{author}{\bibfnamefont{T.~E.} \bibnamefont{Strohmayer}}
  \bibnamefont{and} \bibinfo{author}{\bibfnamefont{A.~L.} \bibnamefont{Watts}},
  \bibinfo{journal}{Astrophys. J.} \textbf{\bibinfo{volume}{653}},
  \bibinfo{pages}{593} (\bibinfo{year}{2006}), \eprint{astro-ph/0608463}.

\bibitem[{\citenamefont{Watts and
  Strohmayer}(2007{\natexlab{a}})}]{Watts:2006ew}
\bibinfo{author}{\bibfnamefont{A.~L.} \bibnamefont{Watts}} \bibnamefont{and}
  \bibinfo{author}{\bibfnamefont{T.~E.} \bibnamefont{Strohmayer}},
  \bibinfo{journal}{Astrophys. Space Sci.} \textbf{\bibinfo{volume}{308}},
  \bibinfo{pages}{625} (\bibinfo{year}{2007}{\natexlab{a}}),
  \eprint{astro-ph/0608476}.

\bibitem[{\citenamefont{Watts and
  Strohmayer}(2007{\natexlab{b}})}]{Watts:2006mr}
\bibinfo{author}{\bibfnamefont{A.~L.} \bibnamefont{Watts}} \bibnamefont{and}
  \bibinfo{author}{\bibfnamefont{T.~E.} \bibnamefont{Strohmayer}},
  \bibinfo{journal}{Adv. Space Res.} \textbf{\bibinfo{volume}{40}},
  \bibinfo{pages}{1446} (\bibinfo{year}{2007}{\natexlab{b}}),
  \eprint{astro-ph/0612252}.

\bibitem[{\citenamefont{Thompson and Duncan}(1995)}]{Thompson:1995gw}
\bibinfo{author}{\bibfnamefont{C.}~\bibnamefont{Thompson}} \bibnamefont{and}
  \bibinfo{author}{\bibfnamefont{R.~C.} \bibnamefont{Duncan}},
  \bibinfo{journal}{Mon. Not. Roy. Astron. Soc.}
  \textbf{\bibinfo{volume}{275}}, \bibinfo{pages}{255} (\bibinfo{year}{1995}).

\bibitem[{\citenamefont{Kouveliotou et~al.}(1998)}]{Kouveliotou:1998ze}
\bibinfo{author}{\bibfnamefont{C.}~\bibnamefont{Kouveliotou}}
  \bibnamefont{et~al.}, \bibinfo{journal}{Nature}
  \textbf{\bibinfo{volume}{393}}, \bibinfo{pages}{235} (\bibinfo{year}{1998}).

\bibitem[{\citenamefont{Kouveliotou et~al.}(2003)\citenamefont{Kouveliotou,
  Duncan, and Thompson}}]{Kouveliotou:2003tb}
\bibinfo{author}{\bibfnamefont{C.}~\bibnamefont{Kouveliotou}},
  \bibinfo{author}{\bibfnamefont{R.~C.} \bibnamefont{Duncan}},
  \bibnamefont{and} \bibinfo{author}{\bibfnamefont{C.}~\bibnamefont{Thompson}},
  \bibinfo{journal}{Sci. Am.} \textbf{\bibinfo{volume}{288N2}},
  \bibinfo{pages}{24} (\bibinfo{year}{2003}).

\bibitem[{\citenamefont{Ruester et~al.}(2006)\citenamefont{Ruester, Hempel, and
  Schaffner-Bielich}}]{Ruester:2005fm}
\bibinfo{author}{\bibfnamefont{S.~B.} \bibnamefont{Ruester}},
  \bibinfo{author}{\bibfnamefont{M.}~\bibnamefont{Hempel}}, \bibnamefont{and}
  \bibinfo{author}{\bibfnamefont{J.}~\bibnamefont{Schaffner-Bielich}},
  \bibinfo{journal}{Phys. Rev.} \textbf{\bibinfo{volume}{C73}},
  \bibinfo{pages}{035804} (\bibinfo{year}{2006}), \eprint{astro-ph/0509325}.

\bibitem[{\citenamefont{{Haensel} et~al.}(1989)\citenamefont{{Haensel},
  {Zdunik}, and {Dobaczewski}}}]{Haensel:1989}
\bibinfo{author}{\bibfnamefont{P.}~\bibnamefont{{Haensel}}},
  \bibinfo{author}{\bibfnamefont{J.~L.} \bibnamefont{{Zdunik}}},
  \bibnamefont{and}
  \bibinfo{author}{\bibfnamefont{J.}~\bibnamefont{{Dobaczewski}}},
  \bibinfo{journal}{Astron. Astrophys.} \textbf{\bibinfo{volume}{222}},
  \bibinfo{pages}{353} (\bibinfo{year}{1989}).

\bibitem[{\citenamefont{{Haensel} and {Pichon}}(1994)}]{Haensel:1994}
\bibinfo{author}{\bibfnamefont{P.}~\bibnamefont{{Haensel}}} \bibnamefont{and}
  \bibinfo{author}{\bibfnamefont{B.}~\bibnamefont{{Pichon}}},
  \bibinfo{journal}{Astron. Astrophys.} \textbf{\bibinfo{volume}{283}},
  \bibinfo{pages}{313} (\bibinfo{year}{1994}), \eprint{arXiv:nucl-th/9310003}.

\bibitem[{\citenamefont{Wigner}(1934)}]{Wigner:1934}
\bibinfo{author}{\bibfnamefont{E.}~\bibnamefont{Wigner}},
  \bibinfo{journal}{Phys. Rev.} \textbf{\bibinfo{volume}{46}},
  \bibinfo{pages}{1002} (\bibinfo{year}{1934}).

\bibitem[{\citenamefont{Wigner}(1938)}]{Wigner:1938}
\bibinfo{author}{\bibfnamefont{E.}~\bibnamefont{Wigner}},
  \bibinfo{journal}{Transactions of the Faraday Society}
  \textbf{\bibinfo{volume}{34}}, \bibinfo{pages}{678} (\bibinfo{year}{1938}).

\bibitem[{\citenamefont{Fetter and Walecka}(1971)}]{Fetter:1971}
\bibinfo{author}{\bibfnamefont{A.~L.} \bibnamefont{Fetter}} \bibnamefont{and}
  \bibinfo{author}{\bibfnamefont{J.~D.} \bibnamefont{Walecka}},
  \emph{\bibinfo{title}{Quantum Theory of Many Particle Systems}}
  (\bibinfo{publisher}{McGraw-Hill, New York}, \bibinfo{year}{1971}).

\bibitem[{\citenamefont{Coldwell-Horsfall and Maradudin}(1960)}]{Coldwell:1960}
\bibinfo{author}{\bibfnamefont{R.~A.} \bibnamefont{Coldwell-Horsfall}}
  \bibnamefont{and} \bibinfo{author}{\bibfnamefont{A.~A.}
  \bibnamefont{Maradudin}}, \bibinfo{journal}{Journal of Mathematical Physics}
  \textbf{\bibinfo{volume}{1}}, \bibinfo{pages}{395} (\bibinfo{year}{1960}).

\bibitem[{\citenamefont{Sholl}(1967)}]{Sholl:1967}
\bibinfo{author}{\bibfnamefont{C.~A.} \bibnamefont{Sholl}},
  \bibinfo{journal}{Proceedings of the Physical Society}
  \textbf{\bibinfo{volume}{92}}, \bibinfo{pages}{434} (\bibinfo{year}{1967}).

\bibitem[{\citenamefont{Audi and Wapstra}(1993)}]{Audi:1993zb}
\bibinfo{author}{\bibfnamefont{G.}~\bibnamefont{Audi}} \bibnamefont{and}
  \bibinfo{author}{\bibfnamefont{A.~H.} \bibnamefont{Wapstra}},
  \bibinfo{journal}{Nucl. Phys.} \textbf{\bibinfo{volume}{A565}},
  \bibinfo{pages}{1} (\bibinfo{year}{1993}).

\bibitem[{\citenamefont{Audi and Wapstra}(1995)}]{Audi:1995dz}
\bibinfo{author}{\bibfnamefont{G.}~\bibnamefont{Audi}} \bibnamefont{and}
  \bibinfo{author}{\bibfnamefont{A.~H.} \bibnamefont{Wapstra}},
  \bibinfo{journal}{Nucl. Phys.} \textbf{\bibinfo{volume}{A595}},
  \bibinfo{pages}{409} (\bibinfo{year}{1995}).

\bibitem[{\citenamefont{Duflo}(1994)}]{Duflo:1994}
\bibinfo{author}{\bibfnamefont{J.}~\bibnamefont{Duflo}},
  \bibinfo{journal}{Nucl. Phys.} \textbf{\bibinfo{volume}{A576}},
  \bibinfo{pages}{29} (\bibinfo{year}{1994}).

\bibitem[{\citenamefont{Zuker}(1994)}]{Zuker:1994}
\bibinfo{author}{\bibfnamefont{A.}~\bibnamefont{Zuker}},
  \bibinfo{journal}{Nucl. Phys.} \textbf{\bibinfo{volume}{A576}},
  \bibinfo{pages}{65} (\bibinfo{year}{1994}).

\bibitem[{\citenamefont{Duflo and Zuker}(1995)}]{Duflo:1995}
\bibinfo{author}{\bibfnamefont{J.}~\bibnamefont{Duflo}} \bibnamefont{and}
  \bibinfo{author}{\bibfnamefont{A.}~\bibnamefont{Zuker}},
  \bibinfo{journal}{Phys. Rev. C} \textbf{\bibinfo{volume}{52}},
  \bibinfo{pages}{R23} (\bibinfo{year}{1995}).

\bibitem[{\citenamefont{Moller et~al.}(1995)\citenamefont{Moller, Nix, Myers,
  and Swiatecki}}]{Moller:1993ed}
\bibinfo{author}{\bibfnamefont{P.}~\bibnamefont{Moller}},
  \bibinfo{author}{\bibfnamefont{J.~R.} \bibnamefont{Nix}},
  \bibinfo{author}{\bibfnamefont{W.~D.} \bibnamefont{Myers}}, \bibnamefont{and}
  \bibinfo{author}{\bibfnamefont{W.~J.} \bibnamefont{Swiatecki}},
  \bibinfo{journal}{Atom. Data Nucl. Data Tabl.} \textbf{\bibinfo{volume}{59}},
  \bibinfo{pages}{185} (\bibinfo{year}{1995}), \eprint{nucl-th/9308022}.

\bibitem[{\citenamefont{Moller et~al.}(1996)\citenamefont{Moller, Nix, and
  Kratz}}]{Moller:1997bz}
\bibinfo{author}{\bibfnamefont{P.}~\bibnamefont{Moller}},
  \bibinfo{author}{\bibfnamefont{J.~R.} \bibnamefont{Nix}}, \bibnamefont{and}
  \bibinfo{author}{\bibfnamefont{K.~L.} \bibnamefont{Kratz}},
  \bibinfo{journal}{Atom. Data Nucl. Data Tabl.} \textbf{\bibinfo{volume}{66}},
  \bibinfo{pages}{131} (\bibinfo{year}{1996}).

\bibitem[{\citenamefont{Friedman and Pandharipande}(1981)}]{Friedman:1981qw}
\bibinfo{author}{\bibfnamefont{B.}~\bibnamefont{Friedman}} \bibnamefont{and}
  \bibinfo{author}{\bibfnamefont{V.~R.} \bibnamefont{Pandharipande}},
  \bibinfo{journal}{Nucl. Phys.} \textbf{\bibinfo{volume}{A361}},
  \bibinfo{pages}{502} (\bibinfo{year}{1981}).

\bibitem[{\citenamefont{Todd and Piekarewicz}(2003)}]{Todd:2003xs}
\bibinfo{author}{\bibfnamefont{B.~G.} \bibnamefont{Todd}} \bibnamefont{and}
  \bibinfo{author}{\bibfnamefont{J.}~\bibnamefont{Piekarewicz}},
  \bibinfo{journal}{Phys. Rev.} \textbf{\bibinfo{volume}{C67}},
  \bibinfo{pages}{044317} (\bibinfo{year}{2003}), \eprint{nucl-th/0301092}.

\end{thebibliography}

\end{document}